\documentclass{stacs_proc}
\newtheorem{defn}{Definition}[section]
\newtheorem{alg}{Algorithm}
\long\def\@makealgocaption#1#2{\vskip 2ex \small
  \hbox to \hsize{\parbox[t]{\hsize}{{\bfseries #1.} #2}}}
\newcounter{algorithm}
\def\thealgorithm{\@arabic\c@algorithm}
\def\fps@algorithm{tbp}
\def\ftype@algorithm{4}
\def\ext@algorithm{lof}
\def\fnum@algorithm{Algorithm \thealgorithm}
\def\algorithm{\let\@makecaption\@makealgocaption\@float{algorithm}}

\newcommand{\bproof}{\noindent{\it Proof}}
\newcommand{\eproof}{\hspace*{\fill}\rule{2mm}{2mm}~~~~~\bigskip}
\newenvironment{pf}{\bproof: }{\eproof}

\usepackage{amsmath}
\usepackage{amssymb}
\begin{document}
\title{Factoring Polynomials over Finite Fields using Balance Test}
\author{Chandan Saha}{Chandan Saha}
\address{Department of Computer Science and Engineering, Indian
Institute of Technology, Kanpur}
%   \address{Department of Computer Science and Engineering
%     \newline Indian Institute of Technology Kanpur}
\email{csaha@cse.iitk.ac.in}

\begin{abstract}
We study the problem of factoring univariate polynomials over finite fields. Under the assumption of the Extended Riemann Hypothesis (ERH), Gao \cite{gao01} designed a polynomial time algorithm that fails to factor only if the input polynomial satisfies a strong symmetry property, namely \emph{square balance}. In this paper, we propose an extension of Gao's algorithm that fails only under an even stronger symmetry property. We also show that our property can be used to improve the time complexity of best deterministic algorithms on most input polynomials. The property also yields a new randomized polynomial time algorithm.
\end{abstract}

\maketitle

\stacsheading{2008}{609-620}{Bordeaux}
\firstpageno{609}

%%%----------------------------------------------------------------------
\section{Introduction} \label{sec:intro}
We consider the problem of designing an efficient deterministic algorithm for factoring a univariate polynomial, with coefficients taken from a finite field. The problem reduces in polynomial time to the problem of factoring a monic, square-free and completely splitting polynomial $f(x)$ with coefficients in a prime field $F_p$ (see \cite{bkamp70}, \cite{lidlnied94}). Although there are efficient polynomial time randomized algorithms for factoring $f(x)$ (\cite{bkamp70}, \cite{cz81}, \cite{gs92}, \cite{ksh95}), as yet there is no deterministic polynomial time algorithm even under the assumption of the Extended Riemann Hypothesis (ERH). In this paper we will assume that ERH is true and $\xi_1, \xi_2, \ldots, \xi_n$ are the $n$ distinct roots of the input polynomial $f$,
\begin{equation*}
f(x) = \prod_{i=1}^{n} \left ( x - \xi_i \right ) \text{\hspace{0.1in} where $\xi_i \in F_p$}
\end{equation*}

In 2001, Gao \cite{gao01} gave a deterministic factoring algorithm that fails to find nontrivial factors of $f$ in polynomial time, if $f$ belongs to a restricted class of polynomials, namely \emph{square balanced polynomials}. Motivated by the work of Gao \cite{gao01}, we have defined a proper subclass of square balanced polynomials, namely \emph{cross balanced polynomials}, such that polynomials that are not cross balanced, can be factored deterministically in polynomial time, under the assumption of the ERH.

Our contribution can be summarized as follows. Let $f$ be a monic, square-free and completely splitting polynomial in $F_p[x]$ with $n$ roots $\xi_1, \ldots, \xi_n$. Our factoring algorithm uses an arbitrary (but deterministically chosen) collection of $k = (n \log p)^{O(1)}$ ($n = deg(f)$) small degree auxiliary polynomials $p_1(.), \ldots, p_k(.)$, and from each $p_l(\cdot)$ $(1 \leq l \leq k)$ and $f$ it implicitly constructs a simple $n$-vertex digraph $G_l$ such that, (for $l>1$) $G_l$ is a subgraph (not necessarily a proper subgraph) of $G_{l-1}$. A proper factor of $f$ is efficiently retrieved if any one of the graphs is either not regular, or is regular with in degree and out degree of every vertex less than a chosen constant $c$. This condition of regularity of all the $k$ graphs imposes a \emph{tight symmetry condition} on the roots of $f$, and we point out that this may be exploited to improve the worst case time complexity of the best known deterministic algorithms. Further, we show that if the polynomials $p_l(\cdot)$ $(1 \leq l \leq k)$ are randomly chosen then the symmetry breaks with high probability and our algorithm works in randomized polynomial time. We call the checking of this symmetry condition a \emph{balance test}.

We now present a little more details. Define the sets $\Delta_i$ for $1 \leq i \leq n$ as,
\begin{equation*}
\Delta_i = \{ 1 \leq j \leq n : j \neq i, \sigma((\xi_i - \xi_j)^2) = -(\xi_i - \xi_j)\}
\end{equation*}
where $\sigma$ is the square root algorithm described in \cite{gao01} (see section \ref{subsec:gaoalgo}). %(see section $2.4$  in \cite{gao01}).
The polynomial $f$ is called a \emph{square balanced polynomial} (as in \cite{gao01}) if $\#\Delta_1 = \ldots = \#\Delta_n$. For $l > 1$, define polynomial $f_l$ as,
\begin{equation*}
f_l = \prod_{i = 1}^{n}{\left( x - p_l(\xi_i) \right)}
\end{equation*}
where $p_l(.)$ is an arbitrary but deterministically chosen polynomial with degree bounded by $(n \log p)^{O(1)}$. Further, $p_{l_1}(.) \neq p_{l_2}(.)$ for $l_1 \neq l_2$, and $f_1$ is taken to be $f$ i.e. $p_1(y) = y$. Assume that, for a given $k = (n \log p)^{O(1)}$, for every $l$, $1 \leq l \leq k$, polynomial $f_l = \tilde{f}^{d_l}_l$, where $\tilde{f}_l$ is a square-free and square balanced polynomial and $d_l > 0$. Later, we show that, if $f_l$ is not of the above form then a proper factor of $f$ can be retrieved efficiently. For each polynomial $f_l$, $1 \leq l \leq k$, define the sets $\Delta_i^{(l)}$ for $1 \leq i \leq n$ as,
\begin{equation*}
\Delta_i^{(l)} = \{1 \leq j \leq n : p_l(\xi_i) \neq p_l(\xi_j), \sigma((p_l(\xi_i) - p_l(\xi_j))^2) = -(p_l(\xi_i) - p_l(\xi_j))\}
\end{equation*}
Further, define the sets ${D_i}^{(l)}$ iteratively over $l$ as,
\begin{eqnarray*}
D_i^{(1)} &=& \Delta_i^{(1)} \\
\text{For $l > 1$, } D_i^{(l)} &=& D_i^{(l-1)} \cap \Delta_i^{(l)}\\
\text{If } D_i^{(l)} &=& \phi \text{ for all $i$, $1 \leq i \leq n$, then redefine $D_i^{(l)}$ as $D_i^{(l)}=D_i^{(l-1)}$}.
\end{eqnarray*}
For $ 1 \leq l \leq k$, let $G_l$ be a directed graph with $n$ vertices $v_1, \ldots, v_n$, such that there is an edge from $v_i$ to $v_j$ if and only if $j \in D_i^{(l)}$. Note that, $G_l$ is a subgraph of $G_{l-1}$ for $1 < l \leq k$. Denote the in degree and out degree of a vertex $v_i$ by $indeg(v_i)$ and $outdeg(v_i)$, respectively. We say that the graph $G_l$ is \emph{regular} (or \emph{$t$-regular}) if $indeg(v_1) = outdeg(v_1) = \ldots = indeg(v_n) = outdeg(v_n) = t$. Call $t$ as the \emph{regularity} of $G_l$. The following theorem is proved in this paper.
\begin{thm}\label{thm:crossbalance}
Polynomial $f$ can be factored into nontrivial factors in time $l \cdot (n \log p)^{O(1)}$ if $G_l$ is not regular for some $l$, $1 \leq l \leq k$. Further, if $G_1, \ldots, G_k$ are all regular and for at least $\lceil \log_2 n \rceil$ of the graphs we have $G_l \neq G_{l-1}$ $(1 < l \leq k)$, then $f$ can be factored in $k \cdot (n \log p)^{O(1)}$ time.
\end{thm}
\noindent Note that, $G_1$ is regular if and only if $f$ is square balanced, as $\Delta_i^{(1)} = \Delta_i$, for $1 \leq i \leq n$ and $G_1$ is in fact a regular tournament.\\

Suppose $f(y)$ splits as $f(y) = (y - X) \cdot f'(y)$ in the quotient ring $R = \frac{F_p[x]}{(f)}$ where $X = x \mod f$. Our algorithm iteratively tests graphs $G_1, G_2, \ldots$ so on, to check if any one of them is not regular. If at the $l^{th}$ iteration graph $G_l$ turns out to be not regular, then a proper factor of $f$ is obtained in polynomial time. However, if $G_l$ is regular, then the algorithm returns a nontrivial monic factor $g_l(y)$ of $f'(y)$ with degree equal to the regularity of $G_l$. Moreover, $g_l(y)$ is also a factor of (although may be equal to) $g_{l-1}(y)$, the factor obtained at the $(l - 1)^{th}$ iteration, and it can be ensured that if $g_l(y)$ is a proper factor of $g_{l-1}(y)$ (which happens iff $G_l \neq G_{l-1}$) then $deg(g_l(y)) \leq \frac{1}{2} \cdot deg(g_{l-1}(y))$. Thus, if the graphs repeatedly turn out to be regular (which in itself is a stringent condition) and for at least $\lceil \log_2 n \rceil$ times it happen that $G_l \neq G_{l-1}$, for $1 < l \leq k$, then we obtain a nontrivial linear factor $g(y)$ of $f'(y)$. The element $-g(0)$ defines a nontrivial \emph{endomorphism} in the ring $R$, and by using a result from \cite{evdo94} (Lemma $9$ in \cite{evdo94}) we can find a proper factor of $f$ in polynomial time. Further, if for only $\epsilon \lceil \log_2 n \rceil$ times we get $G_l \neq G_{l-1}$ $(1 < l \leq k )$ for some $\epsilon$, $0 < \epsilon \leq 1$, then we obtain a nontrivial factor $g(y)$ of $f'(y)$ with degree at most $\frac{n^{1 - \epsilon}}{2}$. Now if we apply Evdokimov's algorithm (\cite{evdo94}) on $g(y)$ (instead of $f'(y)$), we can get a proper factor of $f$ in time $(n^{\frac{(1 - \epsilon)^2}{2}\log n + \epsilon + c_1} \log p)^{c_2}$ ($c_1$ and $c_2$ are constants). For most polynomials $\epsilon > 0$ (i.e. at least about $\frac{1}{\log n}$) and this gives an improvement over the time complexity of $(n^{\frac{1}{2}\log n + c_1} \log p)^{c_2}$ in \cite{evdo94} ($c_1$, $c_2$ are the same constants).

Assuming $n << p$, all the best known deterministic algorithms (e.g. \cite{evdo94}, \cite{CH00}) use computations in rings with large dimensions over $F_p$ to get smaller degree factors of $f'(y)$. Unlike these approaches, the balance test is an attempt to exploit an asymmetry among the roots of the input polynomial to obtain smaller degree factors of $f'(y)$ without carrying out computations in rings with large dimensions over $F_p$. This attribute of our approach yields a better time complexity for most polynomials in a way as discussed in the previous paragraph.

It is sufficient to choose the auxiliary polynomials $p_l(y)$, $1 < l \leq k$, in such a way that the graphs, if regular, are not all the same for too long, if their regularities are large. An efficient and deterministic construction of such auxiliary polynomials will immediately imply that factorization of univariate polynomials over finite fields can be done in deterministic polynomial time under ERH. In this paper we assume that the auxiliary polynomials are arbitrary but deterministically chosen polynomials with degree bounded by $(n \log p)^{O(1)}$. For example, one possibility is to choose $p_l(y) = y^l$ for $1 \leq l \leq k$. (In fact, Gao \cite{gao01} used this choice of auxiliary polynomials to define a restricted class of square balanced polynomials called \emph{super square balanced polynomials}.) We show that, if random choices of auxiliary polynomials are allowed then our algorithm works in randomized polynomial time. For the graphs to be all regular and equal, the roots of $f$ must satisfy a tight symmetry condition (given by equal sizes of all the sets $D_i^{(l)}$, for $1 \leq i \leq n$ and $1 \leq l \leq k$) and it is only then that our algorithm fails to factor $f$.
\begin{defn}
A polynomial $f$ is called \emph{$k$-cross balanced}, for $k > 0$, if for every $l$, $1 \leq l \leq k$, polynomial $f_l = \tilde{f}^{d_l}_l$, where $\tilde{f}_l$ is a square-free, square balanced polynomial with $d_l > 0$, and graph $G_l$ is regular.
\end{defn}
\noindent It follows from the definition that, $1$-cross balanced polynomials form the class of square balanced polynomials. Let $k = (n \log p)^{O(1)}$ be some fixed polynomial in $n$ and $\log p$. A polynomial $f$ is called \emph{cross balanced} if it is $k$-cross balanced and regularity of graph $G_k$ is greater than a fixed constant $c$. From Theorem \ref{thm:crossbalance} and \cite{evdo94} it follows that, polynomials that are not cross balanced can be factored deterministically in polynomial time.
\section{Preliminaries} \label{sec:background}
Assume that $f$ is a monic, square-free and completely splitting polynomial over $F_p$ and $R = \frac{F_p[x]}{(f)}$ is the quotient ring consisting of all polynomials modulo $f$.
\subsection{Primitive Idempotents}
Elements $\chi_1, \ldots, \chi_n$ of the ring $R$ are called the \emph{primitive idempotents} of $R$ if, $\sum_{i=1}^{n}{\chi_i} = 1$ and for $1 \leq i, j \leq n$, $\chi_i \cdot \chi_j = \chi_i$ if $i = j$ and $0$ otherwise. By Chinese Remaindering theorem, $R \cong F_p \oplus \ldots \oplus F_p$ ($n$ times), such that every element in $R$ can be uniquely represented by an $n$-tuple of elements in $F_p$. Addition and multiplication between two elements in $R$ can viewed as componentwise addition and multiplication of the $n$-tuples. Any element $\alpha = (a_1, \ldots, a_n) \in R$ can be equated as, $\alpha = \sum_{i = 1}^{n}{a_i \chi_i}$ where $a_i \in F_p$. Let $g(y)$ be a polynomial in $R[y]$ given by,
\begin{eqnarray*}
g(y) &=& \sum_{i=0}^{m}{\gamma_i y^i} \text{\hspace{0.1in} where $\gamma_i \in R$ and}\\
\gamma_i &=& \sum_{j=1}^{n}{g_{ij} \chi_j} \text{\hspace{0.1in} where $g_{ij} \in F_p$ for $0 \leq i \leq m$ and $1 \leq j \leq n$.}
\end{eqnarray*}
Then $g(y)$ can be alternatively represented as,
\begin{equation*}
g(y) = \sum_{j=1}^{n}{g_j(y) \chi_j} \text{ where } g_j(y) = \sum_{i=0}^{m}{g_{ij} y^i} \in F_p[y] \text{ for $1 \leq j \leq n$}.
\end{equation*}
The usefulness of this representation is that, operations on polynomials in $R[y]$ (multiplication, gcd etc.) can be viewed as componentwise operations on polynomials in $F_p[y]$.
\subsection{Characteristic Polynomial}
Consider an element $\alpha = \sum_{i = 1}^{n}{a_i \chi_i} \in R$ where $a_i \in F_p$, $1 \leq i \leq n$. The element $\alpha$ defines a linear transformation on the vector space $R$ (over $F_p$), mapping an element $\beta \in R$ to $\alpha \beta \in R$. The characteristic polynomial of $\alpha$ (viewed as a linear transformation) is independent of the choice of basis and is equal to
\begin{equation*}
c_\alpha(y) = \prod_{i = 1}^{n}{(y - a_i)},
\end{equation*}
In order to construct $c_\alpha$ one can use $1, X, X^2, \ldots, X^{n-1}$ as the basis in $R$ and form the matrix $(m_{ij})$ where $\alpha \cdot X^{j-1} = \sum_{i=1}^{n}{m_{ij} X^{i-1}}$, $m_{ij} \in F_p$, $1 \leq i, j \leq n$. Then $c_\alpha$ can be constructed by evaluating $det(y \cdot I - (m_{ij}))$ at $n$ distinct values of $y$ and solving for the $n$ coefficients of $c_\alpha$ using linear algebra. The process takes only polynomial time. The notion of characteristic polynomial extends even to higher dimensional algebras over $F_p$.
\subsection{GCD of Polynomials}
Let $g(y) = \sum_{i=1}^{n}{g_i(y)\chi_i}$ and $h(y) = \sum_{i=1}^{n}{h_i(y)\chi_i}$ be two polynomials in $R[y]$, where $g_i, h_i \in F_p[y]$ for $1 \leq i \leq n$ . Then, \emph{gcd} of $g$ and $f$ is defined as,
\begin{equation*}
gcd(g, f) = \sum_{i = 1}^{n}{gcd(g_i,h_i) \chi_i}
\end{equation*}
We note that, the concept of \emph{gcd} of polynomials does not make sense in general over any arbitrary algebra. However, the fact that $R$ is a \emph{completely splitting semisimple algebra} over $F_p$ allows us to work component-wise over $F_p$ and this makes the notion of \emph{gcd} meaningful in the context. The following lemma was shown by Gao \cite{gao01}.
\begin{lem}{\cite{gao01}}
Given two polynomials $g, h \in R[y]$, $gcd(g, h)$ can be computed in time polynomial in the degrees of the polynomials, $n$ and $\log p$.
\end{lem}
\subsection{Gao's Algorithm} \label{subsec:gaoalgo}
Let $R = \frac{F_p[x]}{(f)} = F_p[X]$ where $X = x \mod f$ and suppose that $f(y)$ splits in $R$ as, $f(y) = (y - X)f'(y)$. Define quotient ring $S$ as, $S = \frac{R[y]}{(f')} = R[Y]$ where $Y = y \mod f'$. $S$ is an elementary algebra over $F_p$ with dimension $n' = n(n-1)$. Gao \cite{gao01} described an algorithm $\sigma$ for taking square root of an element in $S$. If $p-1 = 2^e w$ where $e \geq 1$ and $w$ is odd, and $\eta$ is a primitive $2^e$-th root of unity, then $\sigma$ has the following properties:
\begin{enumerate}
\item Let $\mu_1, \ldots, \mu_{n'}$ be primitive idempotents in $S$ and $\alpha = \sum_{i = 1}^{n'}{a_i \mu_i} \in S$ where $a_i \in F_p$. Then, $\sigma (\alpha) = \sum_{i = 1}^{n'}{\sigma(a_i) \mu_i}$.
\item Let $a = \eta^u \theta$ where $\theta \in F_p$ with $\theta^w = 1$ and $0 \leq u < 2^e$. Then $\sigma(a^2) = a$ iff $u < 2^{e-1}$.
\end{enumerate}
When $p = 3 \mod 4$, $\eta = -1$ and property $2$ implies that $\sigma(a^2) = a$ for $a\in F_p$ iff $a$ is a quadratic residue in $F_p$.
\begin{alg} \label{alg:squarebalance}
\cite{gao01} \\
Input: A polynomial $f \in F_p[x]$. \\
Output: A proper factor of $f$ or output that ``$f$ is square balanced". \\
1. Form $X$, $Y$, $R$, $S$ as before. \\
2. Compute $C = \frac{1}{2}(X + Y + \sigma((X - Y)^2)) \in S$. \\
3. Compute the characteristic polynomial $c(y)$ of $C$ over $R$. \\
4. Decompose $c(y)$ as $c(y) = h(y)(y - X)^t$, where $t$ is the largest possible. \\
5. If $h(X)$ is a zero divisor in $R$ then find a proper factor of $f$, otherwise output that ``$f$ is square balanced".
\end{alg}
It was shown in \cite{gao01} that Algorithm \ref{alg:squarebalance} fails to find a proper factor of $f$ if and only if $f$ is square balanced. Moreover, it follows from the analysis in \cite{gao01} (see Theorem $3.1$ in \cite{gao01}) that, when $f$ is square balanced the polynomial $h(y)$ takes the form,
\begin{equation*}
h(y) = \sum_{i = 1}^{n}{\left[ \prod_{j \in \Delta_i}{(y - \xi_j)}\right]\chi_i}
\end{equation*}
where $\Delta_i = \{ j : j \neq i, \sigma((\xi_i - \xi_j)^2) = -(\xi_i - \xi_j) \}$ and $\# \Delta_i = \frac{n-1}{2}$ for all $i$, $1 \leq i \leq n$.
\section{Our Algorithm and Analysis} \label{sec:novelbalance}
In this section, we describe our algorithm for factoring polynomial $f$. We show that the algorithm fails to factor $f$ in $k \cdot (n \log p)^{O(1)}$ time if and only if $f$ is $k$-cross balanced and regularity of $G_k$ is greater than $c$. The algorithm involves $k$ polynomials, $f = f_1, \ldots, f_k$, where polynomial $f_l$, $1 < l \leq k$, is defined as,
\begin{equation*}
f_l = \prod_{i = 1}^{n}{\left(x - p_l(\xi_i)\right)}
\end{equation*}
where $p_l(.)$ is an arbitrary but deterministically fixed polynomial with degree bounded by $(n \log p)^{O(1)}$ and $p_{l_1}(.) \neq p_{l_2}(.)$ for $l_1 \neq l_2$. The polynomial $f_l$ can be constructed in polynomial time by considering the element $p_l(X)$ in $R = \frac{F_p[x]}{(f)} = F_p[X]$, where $X = x \mod f$, and then computing its characteristic polynomial over $F_p$.
\begin{lem} \label{lem:propfact}
If $f_l$ is not of the form $f_l = \tilde{f_l}^{d_l}$, where $\tilde{f_l}$ is a square-free, square balanced polynomial and $d_l > 0$, then a proper factor of $f$ can be retrieved in polynomial time.
\end{lem}
\begin{pf}
By definition, $f_l = \prod_{i = 1}^{n}{\left(x - p_l(\xi_i)\right)}$. Define the sets $E_i$, for $1 \leq i \leq n$, as $E_i = \{1 \leq j \leq n : p_l(\xi_j) = p_l(\xi_i) \}$. Consider the following \emph{gcd} in the ring $R[y]$,
\begin{equation*}
g(y) = gcd \left(p_l(y) - p_l(X), f(y) \right) = \sum_{i = 1}^{n}{\left[ \prod_{j \in E_i}{\left(y - \xi_j \right)} \right] \chi_i}
\end{equation*}
The leading coefficient of $g(y)$ is a zero-divisor in $R$, unless $\# E_1 = \ldots = \# E_n = d_l$ (say). Therefore, we can assume that,
\begin{eqnarray*}
f_l &=& \prod_{j = 1}^{m_l}{ \left( x - p_l(\xi_{s_j})\right)^{d_l}} \text{\hspace{0.1in} where $p_l(\xi_{s_1}), \ldots, p_l(\xi_{s_{m_l}})$ are all distinct and $m_l = \frac{n}{d_l}$} \\
&=& \tilde{f_l}^{d_l} \text{\hspace{0.1in} where $\tilde{f_l} = \prod_{j = 1}^{m_l}{ \left( x - p_l(\xi_{s_j})\right)}$ is square-free.}
\end{eqnarray*}
If polynomial $\tilde{f_l}$ (obtained by square-freeing $f_l$) is not square balanced then a proper factor $\tilde{g_l}$ of $\tilde{f_l}$ is returned by Algorithm \ref{alg:squarebalance}. But then,
\begin{equation*}
gcd \left( \tilde{g_l}(p_l(x)), f(x)\right) = \prod_{j: \tilde{g_l}(p_l(\xi_j)) = 0}{ \left( x - \xi_j\right)}
\end{equation*}
is a proper factor of $f$.
\end{pf}

\noindent Algorithm \ref{alg:squarebalance} works with $\tilde{f_l} = \prod_{j = 1}^{m_l}{ \left( x - p_l(\xi_{s_j})\right)}$ as the input polynomial where $p_l(\xi_{s_j})$'s are distinct and $m_l = \frac{n}{d_l}$, and returns a polynomial $h_l(y)$ such that,
\begin{equation} \label{eqn:polyh}
h_l(y) = \sum_{j = 1}^{m_l}{\left[ \prod_{r \in \tilde{\Delta}_j^{(l)}}(y - p_l(\xi_{s_r}))\right] \chi_j^{(l)}}
\end{equation}
where $\chi_j^{(l)}$'s are the primitive idempotents of the ring $R_l = \frac{F_p[x]}{(\tilde{f_l})}$,
\begin{equation*}
\tilde{\Delta}_j^{(l)} = \{ 1 \leq r \leq m_l: r \neq j, \sigma((p_l(\xi_{s_j}) - p_l(\xi_{s_r}))^2) = -(p_l(\xi_{s_j}) - p_l(\xi_{s_r})) \}
\end{equation*}
and $\# \tilde{\Delta}_j^{(l)} = \frac{m_l - 1}{2}$ for $1 \leq j \leq m_l$. Assume that $p > n^2$ and $n$ is odd, as even degree polynomials can be factored in polynomial time. In the following algorithm, parameter $k$ is taken to be a fixed polynomial in $n$ and $\log p$ and $c$ is a fixed constant.
\begin{alg} \label{alg:crossbalance}
Cross Balance \\
Input: A polynomial $f \in F_p[x]$ of odd degree $n$. \\
Output: A proper factor of $f$ or ``Failure".
\begin{itemize}
\item  Choose $k-1$ distinct polynomials $p_2(y), \ldots, p_k(y)$ with degree greater than unity and bounded by a polynomial in $n$ and $\log p $. (We can use any arbitrary, efficient mechanism to deterministically choose the polynomials.) Take $p_1(y) = y$.
\item \textbf{for $l = 1$ to $k$ do}
\begin{enumerate}
\item[][Steps (1) - (2): Constructing polynomial $f_l$ and checking if $f$ can be factored using Lemma \ref{lem:propfact}.] \\
\item (\emph{Construct polynomial $f_l$}) Compute the characteristic polynomial, $c_{\alpha}(x)$, of element $\alpha = p_l(X) \in R$, over $F_p$. Then $f_l = c_{\alpha}(x)$.
\item (\emph{Check if $f$ can be factored}) Check if $f_l$ is of the form $f_l = \tilde{f_l}^{d_l}$, where $\tilde{f_l}$ is a square-free, square balanced polynomial and $d_l > 0$. If not, then find a proper factor of $f$ as in Lemma \ref{lem:propfact}. \\
\item[] [Steps (3) - (6): Constructing graph $G_l$ implicitly.] \\
\item (\emph{Obtain the required polynomial from Algorithm $1$}) Else, $\tilde{f_l}$ is square balanced and Algorithm \ref{alg:squarebalance} returns a polynomial $h_l(y) = y^t + \alpha_1 y^{t-1} + \ldots + \alpha_t$ (as in equation \ref{eqn:polyh}), where $t = \frac{m_l - 1}{2}$ and $\alpha_u \in R_l$ for $1 \leq u \leq t$.
\item (\emph{Change to a common ring so that gcd is feasible}) Each $\alpha_u \in R_l$ is a polynomial $\alpha_u(x) \in F_p[x]$ of degree less than $m_l$. Compute $\alpha'_u$ as, $\alpha'_u = \alpha_u(p_l(x)) \mod f$, for $1 \leq u \leq t$, and construct the polynomial $h'_l(y) = y^t + \alpha'_1 y^{t - 1} + \ldots + \alpha'_t \in R[y]$.
\item (\emph{Construct graph $G_l$ implicitly}) If $l = 1$ then assign $g_l(y) = h'_l(y) \in R[y]$ and continue the loop with the next value of $l$. Else, construct the polynomial $h'_l(p_l(y))$ by replacing $y$ by $p_l(y)$ in $h_l(y)$ and compute $g_l(y)$ as,
\begin{equation*}
g_l(y) = gcd(g_{l - 1}(y), h'_l(p_l(y))) \in R[y].
\end{equation*}
\item (\emph{Check if $G_l$ is a null graph}) Let $g_l(y) = \beta_{t'} y^{t'} + \ldots + \beta_0$, where $t'$ is the degree of $g_l(y)$ and $\beta_u \in R$ for $0 \leq u \leq t'$. If $t' = 0$ then make $g_l(y) = g_{l - 1}(y)$ and continue the loop with the next value of $l$. \\
\item[] [Steps (7) - (8): Checking for equal out degrees of the vertices of graph $G_l$.] \\
\item (\emph{Check if out degrees are equal}) Else, $t' > 0$. If $\beta_{t'}$ is a zero divisor in $R$, construct a proper factor of $f$ from $\beta_{t'}$ and stop.
\item (\emph{Factor if out degrees are small}) Else, if $t' \leq c$ then use Evdokimov's algorithm \cite{evdo94} on $g_l(y)$ to find a proper factor of $f$ in $(n \log p)^{O(1)}$ time. \\
\item[] [Steps (9) - (11): Checking for equal in degrees of the vertices of graph $G_l$.] \\
\item (\emph{Obtain the values of a nice polynomial at multiple points}) If $t' > c$, evaluate $g_l(y) \in R[y]$ at $n \cdot t'$ distinct points $y_1, \ldots, y_{nt'}$ taken from $F_p$. Find the characteristic polynomials of elements $g_l(y_1), \ldots, g_l(y_{nt'}) \in R$ over $F_p$ as $c_1(x), \ldots, c_{nt'}(x) \in F_p[x]$, respectively. Collect the terms $c_i(0)$ for $1 \leq i \leq nt'$.
\item (\emph{Construct the nice polynomial from the values}) Construct the polynomial $r(x) = x^{nt'} + r_1x^{nt' - 1} + \ldots + r_{nt'} \in F_p[x]$ such that $r(y_i) = -c_i(0)$ for $1 \leq i \leq nt'$. Solve for $r_i \in F_p$, $1 \leq i \leq nt'$, using linear algebra.
\item (\emph{Check if in degrees are equal}) For $0 \leq i < t'$, if $f^i(x)$ divides $r(x)$ then compute $gcd \left( \frac{r(x)}{f^i(x)}, f(x)\right) \in F_p[x]$. If a proper factor of $f$ is found, stop. Else, continue with the next value of $l$. \\
\end{enumerate}
\textbf{endfor}
\item If a proper factor of $f$ is \emph{not} found in the above \emph{for} loop, return ``Failure".
\end{itemize}
\end{alg}

\begin{thm} \label{thm:mainthm}
Algorithm \ref{alg:crossbalance} fails to find a proper factor $f$ in $k \cdot (n \log p)^{O(1)}$ time if and only if $f$ is $k$-cross balanced and regularity of graph $G_k$ is greater than $c$.
\end{thm}
\begin{pf}
We show that, Algorithm \ref{alg:crossbalance} fails to find a proper factor of $f$ at the $l^{th}$ iteration of the loop iff $f$ is $l$-cross balanced and regularity of $G_l$ is greater than $c$. Recall the definitions of the sets $\Delta_i^{(l)}$ and $D_i^{(l)}$, $1 \leq i \leq n$, from section \ref{sec:intro}. The set $\Delta_i^{(l)}$ is defined as,
\begin{equation*}
\Delta_i^{(l)} = \{1 \leq j \leq n : p_l(\xi_i) \neq p_l(\xi_j), \sigma((p_l(\xi_i) - p_l(\xi_j))^2) = -(p_l(\xi_i) - p_l(\xi_j))\}
\end{equation*}
And set $D_i^{(l)}$ is defined iteratively over $l$ as,
\begin{eqnarray*}
D_i^{(1)} &=& \Delta_i^{(1)} \\
\text{For $l > 1$, } D_i^{(l)} &=& D_i^{(l-1)} \cap \Delta_i^{(l)}\\
\text{If } D_i^{(l)} &=& \phi \text{ for all $i$, $1 \leq i \leq n$, then $D_i^{(l)}$ is redefined as $D_i^{(l)}=D_i^{(l-1)}$}.
\end{eqnarray*}
Graph $G_l$, with $n$ vertices $v_1, \ldots, v_n$, has an edge from $v_i$ to $v_j$ iff $j \in D_i^{(l)}$.

Algorithm \ref{alg:crossbalance} fails at the first iteration $(l = 1)$ if and only if $f$ is square balanced. In this case, $D_i^{(1)} = \Delta_i^{(1)} = \Delta_i$, the polynomial $g_1(y)$ is,
\begin{equation*}
g_1(y) = h(y) = \sum_{i = 1}^{n}{\left[ \prod_{j \in D_i^{(1)}}(y - \xi_j)\right] \chi_i}
\end{equation*}
and $G_1$ is regular with in degree and out degree of a vertex $v_i$ equal to $\# D_i^{(1)} = \# \Delta_i = \frac{n - 1}{2}$. Thus, polynomial $f$ is $1$-cross balanced and $deg(g_1(y)) = \frac{n - 1}{2}$. If Algorithm \ref{alg:crossbalance} fails at the $l^{th}$ iteration, then we can assume that the polynomials $f = \tilde{f}_1, \ldots, \tilde{f_l}$ are square free and square balanced (by Lemma \ref{lem:propfact}).

Suppose that, Algorithm \ref{alg:crossbalance} fails at the $l^{th}$ iteration. Then, $\tilde{f_l} = \prod_{j = 1}^{m_l}{ \left( x - p_l(\xi_{s_j})\right)}$ is square free and square balanced, and Algorithm \ref{alg:squarebalance} returns the polynomial $h_l(y) \in R_l[y]$ such that,
\begin{equation} \label{eqn:polyh2}
h_l(y) = \sum_{j = 1}^{m_l}{\left[ \prod_{r \in \tilde{\Delta}_j^{(l)}}(y - p_l(\xi_{s_r}))\right] \chi_j^{(l)}}
\end{equation}
where $\chi_j^{(l)}$'s are the primitive idempotents of the ring $R_l = \frac{F_p[x]}{(\tilde{f_l})}$ and,
\begin{equation*}
\tilde{\Delta}_j^{(l)} = \{ 1 \leq r \leq m_l: r \neq j, \sigma((p_l(\xi_{s_j}) - p_l(\xi_{s_r}))^2) = -(p_l(\xi_{s_j}) - p_l(\xi_{s_r})) \}
\end{equation*}
Let, $h_l(y) = y^t + \alpha_1y^{t - 1} + \ldots + \alpha_t$, where $t = \frac{m_l - 1}{2}$ and $\alpha_u \in R_l$ for $1 \leq u \leq t$. Each $\alpha_u \in R_l$ is a polynomial $\alpha_u(x) \in F_p[x]$ with degree less than $m_l$ and if $\alpha_u = \sum_{j = 1}^{m_l}{a_{uj} \chi_j^{(l)}}$ for $a_{uj} \in F_p$, then by Chinese Remaindering theorem (and assuming the correspondence between $\chi_j^{(l)}$ and the factor $(x - p_l(\xi_{s_j}))$ of $\tilde{f_l}$) we get,
\begin{eqnarray*}
\alpha_u(x) &=& q(x) (x - p_l(\xi_{s_j})) + a_{uj} \text{\hspace{0.1in} for some polynomial $q(x) \in F_p[x]$}\\
\Rightarrow \alpha_u(p_l(x)) &=& q(p_l(x))(p_l(x) - p_l(\xi_{s_j})) + a_{uj} \\
\Rightarrow \alpha_u(p_l(x)) &=&  a_{uj} \mod (x - \xi) \text{\hspace{0.1in} for every $\xi \in \{\xi_1, \ldots, \xi_n \}$ such that $p_l(\xi) = p_l(\xi_{s_j})$ }
\end{eqnarray*}
Suppose that, for a given $i$ ($1 \leq i \leq n$), $j(i)$ ($1 \leq j(i) \leq m_l$) is a unique index such that, $p_l(\xi_i) = p_l(\xi_{s_{j(i)}})$. Then, the polynomial $\alpha'_u(x) = \alpha_u(p_l(x)) \mod f$ has the following \emph{direct sum (or canonical)} representation in the ring $R$,
\begin{equation*}
\alpha'_u(x) = \sum_{i = 1}^{n}{a_{uj(i)} \chi_i}
\end{equation*}
This implies that the polynomial $h'_l(y) = y^t + \alpha'_1 y^{t - 1} + \ldots + \alpha'_t \in R[y]$ has the \emph{canonical} representation,
\begin{equation} \label{eqn:polyh'}
h'_l(y) = \sum_{i = 1}^{n}{\left[ \prod_{r \in \tilde{\Delta}_{j(i)}^{(l)}}{(y - p_l(\xi_{s_r}))}\right] \chi_i}
\end{equation}
Inductively, assume that $g_{l - 1}(y)$ has the form,
\begin{equation*}
g_{l - 1}(y) = \sum_{i = 1}^{n}{\left[ \prod_{j \in D_i^{(l - 1)}}{(y - \xi_j)}\right] \chi_i}
\end{equation*}
Then,
\begin{eqnarray*}
g_l(y) &=& gcd \left( g_{l - 1}(y), h'_l(p_l(y)) \right) \\
&=& \sum_{i = 1}^{n}{gcd \left(\prod_{j \in D_i^{(l - 1)}}{(y - \xi_j)}, \prod_{r \in \tilde{\Delta}_{j(i)}^{(l)}}{(p_l(y) - p_l(\xi_{s_r}))} \right) \chi_i} \\
&=& \sum_{i = 1}^{n}{\left[ \prod_{j \in D_i^{(l - 1)} \cap \Delta_i^{(l)}}{(y - \xi_j)}\right] \chi_i} \text{\hspace{0.1in} (as $r \in \tilde{\Delta}_{j(i)}^{(l)} \Leftrightarrow s_r \in  \Delta_i^{(l)}$) }
\end{eqnarray*}
Therefore,
\begin{eqnarray*}
g_l(y) &=& \sum_{i = 1}^{n}{\left[ \prod_{j \in D_i^{(l)}}{(y - \xi_j)} \right] \chi_i} \\
&=& \beta_{t'} y^{t'} + \ldots + \beta_0 \text{\hspace{0.1in} (say)}
\end{eqnarray*}
where $t' = max_i \left( \# D_i^{(l)} \right)$ and $\beta_u \in R$ for $1 \leq u \leq t' \leq \frac{n - 1}{2}$. The element $\beta_{t'}$ is not a zero divisor in $R$ if and only if $\# D_1^{(l)} = \ldots = \# D_n^{(l)} = t'$. If $t' \leq c$ then a factor of $f$ can be retrieved from $g_l(y)$ in polynomial time using already known methods (\cite{evdo94}). The condition $\# D_i^{(l)} = t'$ for all $i, 1 \leq i \leq t'$, makes the \emph{out} degree of every vertex in $G_l$ equal to $t'$. However, this may not necessarily imply that the \emph{in} degree of every vertex in $G_l$ is also $t'$.
Checking for identical \emph{in} degrees of the vertices of $G_l$ is handled in steps $(9) - (11)$ of the algorithm. Consider evaluating the polynomial $g_l(y)$ at a point $y_s \in F_p$.
\begin{equation*}
g_l(y_s) = \sum_{i = 1}^{n}{\left[ \prod_{j \in D_i^{(l)}}{(y_s - \xi_j)}\right] \chi_i} \in R
\end{equation*}
The characteristic polynomial of $g_l(y_s)$ over $F_p$ is,
\begin{eqnarray*}
c_s(x) &=& \prod_{i = 1}^{n}{\left( x - \prod_{j \in D_i^{(l)}}{(y_s - \xi_j)}\right) } \\
\Rightarrow -c_s(0) &=& \prod_{j = 1}^{n}{(y_s - \xi_j)^{k_j}} \text{\hspace{0.1in} (since $n$ is odd)}
\end{eqnarray*}
where $k_j$ is the \emph{in} degree of vertex $v_j$ in $G_l$. Let $r(x) = x^{nt'} + r_1 x^{nt' - 1} + \ldots + r_{nt'} \in F_p[x]$ be a polynomial of degree $nt'$, such that,
\begin{equation*}
r(y_s) = - c_s(0) = \prod_{j = 1}^{n}{(y_s - \xi_j)^{k_j}}
\end{equation*}
for $nt'$ distinct points $\{y_s\}_{1 \leq s \leq nt'}$ taken from $F_p$. Since we have assumed that $p > n^2 > \frac{n(n-1)}{2} \geq nt'$, we can solve for the coefficients $r_1, \ldots, r_{nt'}$ using any $nt'$ distinct points from $F_p$. Then,
\begin{equation*}
r(x) = \prod_{j = 1}^{n}{(x - \xi_j)^{k_j}}
\end{equation*}
If $k_j \neq t'$ for some $j$, then there is an $i = min \{ k_1, \ldots, k_n\} < t'$ such that $f^i(x)$ divides $r(x)$ and $gcd \left( \frac{r(x)}{f^i(x)}, f(x)\right)$ yields a nontrivial factor of $f(x)$. This shows that the graph $G_l$ is regular if the algorithm fails at the $l^{th}$ step. Since $deg(g_l(y))$ equals the regularity of $G_l$, hence if the latter quantity is less than $c$ then we can apply Evdokimov's algorithm \cite{evdo94} on $g_l(y)$ and get a non trivial factor of $f$ in polynomial time.
\end{pf}

Let $H_l$ ($1 \leq l \leq k$) be a digraph with $n$ vertices $v_1, \ldots, v_n$ such that there is an edge from $v_i$ to $v_j$ iff $j \in \Delta_i^{(l)}$. Then, graph $G_l = G_{l-1} \cap H_l$ or $G_l = G_{l-1}$ (if $G_{l-1} \cap H_l = \Phi$, where $\Phi$ is the null graph with $n$ vertices but no edge). Here $\cap$ denotes the edge intersection of graphs defined on the same set of vertices. Algorithm \ref{alg:crossbalance} fails to find a proper factor of $f$ in polynomial time if and only if there exists an $l \leq k$ such that $G_l$ is $t$-regular ($t > c$) and $G_l \cap H_j = G_l$ or $\Phi$ for all $j$, $l < j \leq k$. It is therefore important to choose the polynomials $p_j(\cdot)$ in such a way that very quickly we get a graph $H_j$ with $G_l \cap H_j \neq G_l$ or $\Phi$. We say that a polynomial $p_l(\cdot)$ is good if either $H_l$ is not regular or $G_l \neq G_{l-1}$ $(1 < l \leq k)$. We show that, only a few good polynomials are required.
\begin{lem} \label{lem:fewgoodpoly}
Algorithm \ref{alg:crossbalance} (with a slight modification) requires at most $\lceil \log_2 n \rceil$ good auxiliary polynomials to find a proper factor of $f$.
\end{lem}
\begin{pf}
Consider the following modification of Algorithm \ref{alg:crossbalance}. At step $5$ of Algorithm \ref{alg:crossbalance}, for $l > 1$, take $g_l(y)$ to be either $gcd(g_{l-1}(y), h'_l(p_l(y)))$ or $g_{l-1}(y)/gcd(g_{l-1}(y), h'_l(p_l(y)))$, whichever has the smaller nonzero degree. Accordingly, we modify the definition of graph $G_l$. Define the set $\bar{\Delta}_i^{(l)}$ $(1 \leq i \leq n)$ as,
\begin{equation*}
\bar{\Delta}_i^{(l)} = \{1 \leq j \leq n : j \neq i, \sigma((p_l(\xi_i) - p_l(\xi_j))^2) = (p_l(\xi_i) - p_l(\xi_j))\} = \{1 \leq j \leq n : j \neq i\} - \Delta_i^{(l)}
\end{equation*}
and modify the definition of the sets $D_i^{(l)}$ $(1 \leq i \leq n)$ as,
\begin{eqnarray*}
D_i^{(1)} &=& \Delta_i^{(1)} \\
\text{For $l > 1$, } {D_i}^{(l)} &=& {D_i}^{(l-1)} \cap \Delta_i^{(l)} \text{ if } g_l(y) = gcd(g_{l-1}(y), h'_l(p_l(y)))\\
&=& {D_i}^{(l-1)} \cap \bar{\Delta}_i^{(l)} \text{ else if } g_l(y) = g_{l-1}(y) / gcd(g_{l-1}(y), h'_l(p_l(y)))
\end{eqnarray*}
As before, an edge $(v_i,v_j)$ is present in $G_l$ iff $j \in D_i^{(l)}$. This modification ensures that, if $g_l(y) \neq g_{l-1}(y)$ has an invertible leading coefficient (i.e if $g_l(y)$ is monic) then the degree of $g_l(y)$ is at most half the degree of $g_{l-1}(y)$. Hence, for every good choice of polynomial $p_l(\cdot)$ if $G_{l-1}$ and $G_l$ are $t_{l-1}$-regular and $t_l$-regular, respectively, then $t_l \leq \frac{t_{l-1}}{2}$. Therefore, at most $\lceil \log_2 n \rceil$ good choices of polynomials $p_l(\cdot)$ are required by the algorithm.
\end{pf}

Theorem \ref{thm:crossbalance} follows as a corollary to Theorem \ref{thm:mainthm} and Lemma \ref{lem:fewgoodpoly}. As already pointed out in section \ref{sec:intro}, if only $\epsilon \lceil \log_2 n \rceil$ good auxiliary polynomials are available for some $\epsilon$, $0 < \epsilon \leq 1$, then we obtain a nontrivial factor $g(y)$ of $f'(y)$ with degree at most $\frac{n^{1 - \epsilon}}{2}$. If we apply Evdokimov's algorithm on $g(y)$ instead of $f'(y)$, then the maximum dimension of the rings considered is bounded by $n^{\frac{(1 - \epsilon)^2}{2} \log n + \epsilon + O(1)}$ instead of $n^{\frac{\log n}{2} + O(1)}$ (as is the case in \cite{evdo94}). \\

In the following discussion we briefly analyze the performance of Algorithm \ref{alg:crossbalance} based on uniform random choices of the auxiliary polynomials $p_l(.)$ ($1 < l \leq k$). The proofs are omitted.
\begin{lem} \label{lem:probbound1}
If $p = 3 \mod 4$ and $p \geq n^6 2^{2n}$ then about $\frac{(1 + o(1))^n}{{(\frac{\pi}{2}n)}^{\frac{n}{2}}}$ fraction of all completely splitting, square-free polynomials of degree $n$ are square balanced.
\end{lem}
\begin{cor}
If $p = 3 \mod 4$, $p > n^62^{2n}$ and $p_l(y)$ is a uniformly randomly chosen polynomial of degree $(n - 1)$ then the probability that $f_l$ is either not square-free or is a square-free and square balanced polynomial is upper bounded by $\frac{(1 + o(1))^n}{{(\frac{\pi}{2}n)}^{\frac{n}{2}}}$.
\end{cor}
It follows that, for $p = 3 \mod 4$ and $p > n^62^{2n}$, if the auxiliary polynomials $p_l(\cdot)$'s are uniformly randomly chosen then Algorithm \ref{alg:crossbalance} works in randomized polynomial time.
However, the arguments used in the proof of Lemma \ref{lem:probbound1} do not immediately apply to the case $p = 1 \mod 4$. Therefore, we resort to a more straightforward analysis, although in the process we get a slightly weaker probability bound.
\begin{lem} \label{lem:probbound2}
If $G_l$ $(1 \leq l < k)$ is regular and $p_{l+1}(y) \in F_p[y]$ is a uniformly randomly chosen polynomial of degree $(n-1)$ then $G_{l+1} \neq G_l$ with probability at least $1 - \frac{1}{2^{0.9n  - 2}}$.
\end{lem}
\noindent Thus, if polynomials $p_l(y)$, $1 < l \leq \lceil \log_2 n \rceil$, are randomly chosen, then the probability that $f$ is not factored by Algorithm \ref{alg:crossbalance} within $\lceil \log_2 n \rceil$ iterations is less than $\frac{\lceil \log_2 n \rceil}{2^{0.9n - 2}}$.
\section{Conclusion}
In this paper, we have extended the square balance test by Gao \cite{gao01} and showed a direction towards improving the time complexity of the best previously known deterministic factoring algorithms. Using certain auxiliary polynomials, our algorithm attempts to exploit an inherent asymmetry among the roots of the input polynomial $f$ in order to efficiently find a proper factor. The advantage of using auxiliary polynomials is that, unlike \cite{evdo94}, it avoids the need to carry out computations in rings with large dimensions, thereby saving overall computation time to a significant extent. Motivated by the stringent symmetry requirement from the roots of $f$, we pose the following question:
\begin{itemize}
\item Is it possible to construct good auxiliary polynomials in deterministic polynomial time?
\end{itemize}
\noindent An affirmative answer to the question will immediately imply that factoring polynomials over finite fields can be done in deterministic polynomial time under ERH.
\section*{Acknowledgements}
The author would like to thank Manindra Agrawal and Piyush Kurur for many insightful discussions that helped in improving the result. The suggestions from anonymous referees have significantly improved the presentation of this paper. The author is thankful to them.
\bibliographystyle{alpha}

\begin{thebibliography}{vzGS92}

\bibitem[Ber70]{bkamp70}
E.~R. Berlekamp.
\newblock Factoring polynomials over large finite fields.
\newblock {\em Mathematics of Computation}, 24(111):713--735, 1970.

\bibitem[CH00]{CH00}
Qi~Cheng and Ming-Deh~A. Huang.
\newblock Factoring polynominals over finite fields and stable colorings of
  tournaments.
\newblock {\em ANTS}, pages 233--246, 2000.

\bibitem[CZ81]{cz81}
David~G. Cantor and Hans Zassenhaus.
\newblock A new algorithm for factoring polynomials over finite fields.
\newblock {\em Mathematics of Computation}, 36(154):587--592, 1981.

\bibitem[Evd94]{evdo94}
Sergei Evdokimov.
\newblock Factorization of polynominals over finite fields in subexponential
  time under {GRH}.
\newblock {\em ANTS}, pages 209--219, 1994.

\bibitem[Gao01]{gao01}
Shuhong Gao.
\newblock On the deterministic complexity of factoring polynomials.
\newblock {\em Journal of Symbolic Computation}, 31(1--2):19--36, 2001.

\bibitem[KS95]{ksh95}
Erich Kaltofen and Victor Shoup.
\newblock Subquadratic-time factoring of polynomials over finite fields.
\newblock {\em STOC}, pages 398--406, 1995.

\bibitem[LN94]{lidlnied94}
R.~Lidl and H.~Niederreiter.
\newblock Introduction to finite fields and their applications, revised
  edition.
\newblock {\em Cambridge University Press}, 1994.

\bibitem[vzGS92]{gs92}
Joachim von~zur Gathen and Victor Shoup.
\newblock Computing frobenius maps and factoring polynomials.
\newblock {\em Computational Complexity}, 2:187--224, 1992.

\end{thebibliography}

\end{document}